\documentstyle[aps,multicol,epsf]{revtex}

\newcommand{\beq}{\begin{equation}}
\newcommand{\eeq}{\end{equation}}

\tighten
\begin{document}

\title{Scaling exponents for Barkhausen avalanches in
polycrystals and amorphous ferromagnets}

\author{Gianfranco Durin$^1$ and Stefano Zapperi$^{2}$}

\address{$^1$ Istituto Elettrotecnico Nazionale Galileo Ferraris and
        INFM, Corso M. d'Azeglio 42, I-10125 Torino, Italy\\
        $^2$ INFM sezione di Roma 1, Dipartimento di Fisica,
    Universit\`a "La Sapienza", P.le A. Moro 2
        00185 Roma, Italy}
\maketitle
\begin{abstract}
We investigate the scaling properties of the Barkhausen effect,
recording the noise in several soft ferromagnetic materials:
polycrystals with different grain sizes and amorphous alloys.  We
measure the Barkhausen avalanche distributions and determine the
scaling exponents. In the limit of vanishing external field rate,
we can group the samples in two distinct classes, characterized
by exponents $\tau = 1.50 \pm0.05$ or $\tau = 1.27 \pm 0.03$, for
the avalanche size distributions. We interpret these results in
terms of the depinning transition of domain walls and obtain an
expression relating  the cutoff of the distributions to the
demagnetizing factor which is in quantitative agreement with
experiments.
\end{abstract}


\begin{multicols}{2}

The Barkhausen noise is an indirect measure of complex microscopic
magnetization processes and is commonly employed as a tool to
investigate ferromagnetic materials \cite{Bertotti}. In recent
years, the interest for this phenomenon has grown considerably due
to the connections with disordered systems and non-equilibrium
critical phenomena.

Experiments have shown that the histogram of Barkhausen jump
(avalanche) sizes follows a power law distribution
\cite{BER-94,DUR-95a,URB-95b,SPA-96,ZAP-98}, suggesting the
presence of an underlying critical point
\cite{URB-95b,PER-95,CIZ-97}.  This hypothesis implies that the
statistical properties of the noise should be described by
universal scaling laws, with critical exponents that are
independent of the material microstructure, at least for some
class of materials. However, the exact nature of the critical
behavior is still debated.  Several authors
\cite{URB-95b,ZAP-98,CIZ-97,NAR-96} have analyzed the dynamics of
flexible domain walls in random media relating the Barkhausen
exponents to the scaling expected at the depinning transition.
Other theoretical explanations involve a critical point tuned by
the disorder in the the framework of disordered spin models
\cite{PER-95,VIV-95}.

The experiments reported in the literature can hardly resolve the
theoretical issues since the measured exponents span a relatively wide
range \cite{BER-94,DUR-95a,URB-95b,SPA-96,BAH-99} and
it is difficult to confirm whether universality holds.
In particular, there has been no extensive and systematic
measurement of critical exponents in different
materials with homogeneous and controlled experimental
conditions and reliable statistics.
For example, experimental evidence of universality has
recently been reported for acoustic emission avalanches
recorded during martensitic  transformations, by analyzing in detail
the behavior of several alloys with different compositions
\cite{CAR-98}.

The precise dependence of the Barkhausen
characteristic sizes on experimentally tunable parameters is still
a debated question
\cite{URB-95b,ZAP-98,PER-95,CIZ-97,NAR-96,BAH-99} and a complete
agreement between theory,  simulations and experiments is still
lacking.  Understanding this point is crucial in order to link the
material microstructure to the noise properties, or conversely to
use the noise to obtain information on the structure of the
material. In the past, this problem, which has important
technological relevance, has been mainly addressed using
phenomenological models \cite{Bertotti,ALE-90} where the material
properties (i.e. grain size \cite{BER-90b}, internal stresses
\cite{SAB-93}) are accounted for by effective fitting parameters.
The main limit of this approach is that there is no systematic way
to implement the program, without a precise understanding of the
microscopic dynamics.

In this letter, we report experimental data for a large set of
materials suggesting the existence of two distinct
universality classes and show that the results are in
quantitative agreement with the theory of domain wall depinning
transition. We perform Barkhausen noise measurements in six
different ferromagnetic materials under similar experimental
conditions, averaging the distributions over a large number
($\sim 10^5-10^6$) of events, carefully testing the effect of the
magnetic field rate on the exponents. The cutoff of the
distributions is tuned by the demagnetizing factor and defines
two new critical exponents, which we measure for two materials
belonging to the different classes. Using scaling arguments, we
predict values for these exponents that are in good agreement
with experiments.

We record the Barkhausen noise using standard inductive methods,
described in details in Refs. \cite{DUR-95a,SPA-96,ALE-90}. A long
solenoid provides an homogeneous low frequency triangular driving
field, while a secondary pickup around the sample cross section
gets the induced flux. The solenoid is 60 cm long, with a value of
1450 turns/meter, ensuring an homogeneous field up to 55 cm long
samples with peak amplitude of about 150 A/m. The pickup is made
of 50 isolated copper turns, wounded within 1 mm. Such a small
width is required to avoid spurious effects due to demagnetizing
fields. All the measurements are performed only in the central
part of the hysteresis loop around the coercive field, where
domain wall motion is the relevant magnetization mechanism
\cite{Bertotti}. We take special care to reduce excess external
noise during the measurements of avalanches distributions, as the
evaluation of critical exponents is strongly affected by spurious
noise. In this respect, the most appropriate cutoff frequency of
the low pass pre-amplifier filter is chosen in the 3-20 kHz range,
roughly half of the sampling frequency, as usual in noise
measurements.

The analysis of Barkhausen avalanche distribution is performed
following the procedure discussed in Ref.\cite{DUR-95a}. We impose
a reference level for $v_r$ for the signal $v(t)$, chosen above
the background noise. The duration $T$ of the Barkhausen
avalanches is defined as the interval within two successive
intersections of the signal with the $v=v_r$ line. The avalanche
size $s$ is calculated as the integral of the signal between the
same points. We observe that the avalanche distributions follow a
power law  \beq P(s) = s^{-\tau}f(s/s_0),
~~~~P(T)=T^{-\alpha}g(T/T_0), \eeq where $s_0$ and $T_0$ indicate
the position of the cutoff to the power law behavior. The critical
exponents result to be independent of the reference level for a
reasonable range of $v_r$ \cite{DUR-95a}.

We employ several different soft magnetic materials, both
polycrystalline and amorphous: an Fe-Si 7.8 wt.\% strip (30 cm
$\times$ 0.5 cm $\times$ 60 $\mu$m) produced by plan flow
casting, annealed several times around 950$^\circ$C to obtain grains
of average dimension of 25 $\mu$m; two strips of Fe-Si 6.5 wt.\%
(30 cm $\times$ 0.5 cm $\times$ 45 $\mu$m), one annealed for 2h at
1200$^\circ$C, with grains of 160 $\mu$m, and the other annealed for
2h at 1050$^\circ$C, with grains of 35 $\mu$m \cite{APP-98}. The
amorphous samples have composition of the type
Fe$_{x}$Co$_{85-x}$B$_{15}$ and we employ
Fe$_{21}$Co$_{64}$B$_{15}$ as cast (20 cm $\times$ 1 cm $\times$
22 $\mu$m), Fe$_{64}$Co$_{21}$B$_{15}$ as cast (28 cm $\times$ 1
cm $\times$ 23 $\mu$m). With these highly magnetostrictive alloys
($\lambda_s \sim 30-50 \times 10^{-6}$) a tensile stress of
$\sigma \sim 100$ MPa is applied during the measurement. The
applied stress is found to enhance the signal-noise ratio,
reducing biases in the distributions, but does not change the
exponents \cite{DUR-99}. A partially crystallized
Fe$_{64}$Co$_{21}$B$_{15}$ (22 cm $\times$ 1 cm $\times$ 23
$\mu$m) is also employed, with annealing for 30 min at 350$^\circ$C
and then for 4h at 300$^\circ$C under an applied tensile stress of 500
MPa. This induces the formation of $\alpha$-Fe crystals of about
50 nm, with a crystal fraction of $\sim 5\%$ \cite{BAS-96}.

In Fig.~1a we show the avalanche size distribution, obtained for
the smallest available magnetic field rates ($f$ = 3-5 mHz). We
clearly see that the data can be grouped in two universality
classes with $\tau = 1.50 \pm 0.05$ and $\tau = 1.27 \pm 0.03$.
The first class includes all the Si-Fe polycrystals and the
partially crystallized amorphous alloy, while the amorphous alloys
under stress belong to the second class. For the materials in the
first class, we observed a linear decrease of the exponents on the
frequency $f$ of the external magnetic field, in agreement with
earlier findings \cite{DUR-95a}. The material in the second class
do not show any noticeable dependence of the exponents on the
field rate. We note that $\tau \simeq 1.3$, independent of the
frequency, was previously measured in Perminvar \cite{URB-95b}.
Next, we measure the distribution of avalanche durations (see
Fig.~1b) and find $\alpha=2.0 \pm 0.2$ and $\alpha=1.5 \pm 0.1$
for the two classes, with a quite large error bar due to the
limited range of scaling, and the presence of unavoidable excess
external noise at low durations. Also in this case, $\alpha$
decreases linearly with $f$ for the materials belonging to the
first class.

The scaling of the cutoff of Barkhausen avalanche distributions
has been the object of an intense debate in the literature
\cite{ZAP-98,PER-95,CIZ-97,NAR-96,BAH-99}. In Ref.~\cite{CIZ-97}
the control parameter was identified with the demagnetizing factor
$k$. We thus measure the Barkhausen avalanche distributions
varying $k$, using samples of different aspect ratios. In
particular, we use the same sample and cut it progressively in
shorter pieces, recording the noise always in the same region,
whose size is limited by the pickup coil width. In this way only
$k$ is varied, while stress, internal disorder and system size are
kept constant. The demagnetizing factor is estimated as $k=
1/\mu_c-1/\mu_i$ where $\mu_c$ is the linear permeability around
the coercive field and $\mu_i$ is the intrinsic permeability (i.e.
in an infinite strip) estimated using a magnetic
yoke~\cite{note_k}.

We perform the measurements on the Fe-Si 6.5 wt.\% 1200$^\circ$C (with
lengths spanning from 28 to 10 cm) and the
Fe$_{21}$Co$_{64}$B$_{15}$ (from 27 to 8 cm) under constant
tensile stress. In Fig.~2 we report the avalanche size
distribution for different $k$ for Fe$_{21}$Co$_{64}$B$_{15}$ in
order to show the increase of the cutoff. Data collapse analysis
yields $s_0\sim k^{-1/\sigma_k}$ with $1/\sigma_k \simeq 0.78$
(see the inset of Fig.~2). Similarly the duration distribution
cutoff scales as $s_0\sim k^{-\Delta_k}$ with $\Delta_k \simeq
0.4$. In the case of Fe-Si, the analysis is complicated by the
frequency dependence of the exponents, therefore we fit the cutoff
for different values of $f$ and extrapolate the results for $f\to
0$. The results for $s_0$ and $T_0$ for both materials are
reported in Fig.~3 and Table \ref{tab}.

To interpret the experimental results we use the model of domain
wall depinning discussed in Ref.~\cite{ZAP-98,CIZ-97}. A single
$180^{\circ}$ domain wall is described by its position
$h(\vec{r})$, dividing two regions of opposite magnetization
directed along the $x$ axis. The total energy for a given
configuration is the sum of different contributions due to
magnetostatic, ferromagnetic and magneto-crystalline interactions,
and gives rise to the following equation of motion
\cite{ZAP-98,CIZ-97}
\[
\Gamma\frac{\partial
 h(\vec{r},t)}{\partial t}= 2\mu_0 M_s H
-k\int d^2r^\prime h(\vec{r}^{\;\prime},t)+
\gamma_w \nabla^2h(\vec{r},t)\]
\beq
+ \int d^2r^\prime J(\vec{r}-\vec{r}^{\;\prime})(h(\vec{r
}^{\;\prime})-h(\vec{r})) +\eta(\vec{r},h), \label{eqm}
\eeq
where $\Gamma$ is the viscosity, $M_s$ is the saturation magnetization,
$H$ is the applied field, $k$ is the demagnetizing factor,
$\gamma_{w}$ is the surface
tension of the wall, $J$ is the kernel
due to dipolar interactions given by
\beq
J(\vec{r}-\vec{r}^{\;\prime})=
\frac{\mu_0M_s^2}{2\pi|\vec{r}-\vec{r}^{\;\prime}|^3}\left(1+
\frac{3(x-x^\prime)^2}{|\vec{r}-\vec{r}^{\;\prime}|^2}\right),
\label{eq:ker} \eeq
and $\eta(\vec{r},h)$ is a Gaussian
uncorrelated random field taking into account all the possible
effects of dislocations, residual stress and non-magnetic
inclusions.

The critical behavior of Eq.~\ref{eqm} has been understood using
renormalization group methods \cite{NAT-92,NAR-93,LES-97}, which
show that at large length scales the critical exponents take
mean-field values \cite{ZAP-98,CIZ-97}. This result is due to the
linear dependence on the momentum of the interaction kernel
(Eq.~\ref{eq:ker}) in Fourier space \cite{ERT-94}. In general, if
we consider an interface whose interaction kernel in momentum
space scales as $J(q)=J_0|q|^\mu$, the upper critical dimension
is given by $d_c=2\mu$ and the values of the exponents depend on
$\mu$ (see Table \ref{tab}). In particular, Eq.~\ref{eqm} yields
$\tau = 3/2$ and $\alpha =2$ (i.e. $\mu =1$), or $\tau \simeq
1.27$ and $\alpha \simeq 1.5$ if dipolar interactions are
neglected (i.e. $\mu=2$) \cite{ZAP-98}. Numerical simulations
confirm the linear dependence of the exponents on the driving frequency
$f$ for $\mu=1$, and no dependence for $\mu=2$ \cite{DUR-99}.

The experimental results are in perfect agreement with the values
predicted using Eq.~\ref{eqm} and suggest that the dipolar interactions
are stronger than surface tension effects in polycrystals, or
whenever small grains are present, while in amorphous alloys under
stress the surface tension is much stronger. Magnetostriction can
be one of the sources of this effect, since the domain wall
surface tension increases with stress $\sigma$ as $\gamma_w \sim
\sqrt{K_0+3/2\lambda_s\sigma}$, where $K_0$ is the zero applied
stress anisotropy and $\lambda_s$ is the saturation
magnetostriction \cite{Bertotti,DUR-99}. Micromagnetic
calculations for these particular materials are needed to have a
final confirmation of this effect.

We derive the dependence of the cutoff on $k$ from Eq.~\ref{eqm}
noting that the demagnetizing field acts as a restoring force for
the interface motion and is responsible for the cutoff in the
avalanche sizes. The interface can not jump over distances larger
than $\xi$, defined as the length for which the interaction term
$J_0|q|^\mu$ is overcome by the restoring force (i.e.
$J_0h\xi^{-\mu} \sim k\xi^d h$) which implies \cite{Nota} \beq \xi
\sim (k/J_0)^{-\nu_k} ~~~~~~\nu_k = 1/(\mu+d). \eeq The avalanche
size and duration distributions cutoff can be obtained using the
scaling relations reported in Ref.~\cite{ZAP-98}:
\beq
s_0 \sim D(k/J_0)^{-1/\sigma_k},~~~~~~1/\sigma_k = \nu_k (d+\zeta)
\label{eq:s0}
\eeq
and similarly
\beq
T_0\sim D (k/J_0)^{-\Delta_k},~~~~~~~\Delta_k = z\nu_k, \label{eq:T0}
\eeq
where $D\equiv\sqrt{\langle\eta^2\rangle}$ denotes the typical
fluctuation of the disorder. The dynamic exponent $z$ and the
interface roughness exponent $\zeta$ define the spatio-temporal
scaling of the domain wall width $\sqrt{\langle h^2\rangle-\langle
h\rangle^2} = \xi^\zeta~F(t/\xi^z)$.

Inserting in Eqs.~(\ref{eq:s0}-\ref{eq:T0}) the renormalization
group results \cite{NAT-92,NAR-93,LES-97,ERT-94} $\zeta = (2\mu-d)/3$ and
$z=\mu-(4\mu-2d)/9$, we obtain $1/\sigma_k = 2/3$ and
$\Delta_k=(\mu-(4\mu-2d)/9)/(\mu+d)$. We have performed extensive
numerical simulations in $d=2$ using the model described in
Ref.~\cite{ZAP-98} in order to test these results (see
Table \ref{tab}). Our results also
agree with earlier numerical simulations in $d=1$ for $1<\mu<2$ where
$s_0\sim k^{-0.65}$ independent of $\mu$ \cite{TAN-98} and with
the result reported in Ref.~\cite{BAH-99}
(i.e. $s_0 \sim L^{1.4}$) obtained using $k \equiv 1/L^2$.

The scaling of the cutoff is different for a {\it local}
demagnetizing field $-k h(x,t)$ \cite{URB-95b}, which yields
$\nu_k=1/\mu$ \cite{NAR-96}. However, a non-local term is more
appropriate to describe the demagnetizing field, due to the
long-range of magnetostatic interactions \cite{ZAP-98}, as it is
confirmed by the agreement between experiments and theory.

In conclusions, our experiments suggest that
the Barkhausen effect can be described by universal
scaling functions and that materials can be classified
in different universality classes. The theory of
interface depinning can be used to obtain a
quantitative explanation of the experiments, providing
a natural framework to understand the properties
of soft magnetic materials.

\begin{table}
\begin{tabular}{|c|c|c|c|c|}
\hline
  $\mu = 1$  &    $\tau$     & $\alpha$ & $1/\sigma_k$  &   $\Delta_k$
\\
\hline
    RG     &      3/2      &   2      &      2/3      &      1/3
\\
\hline
    Sim     &      1.50$\pm$0.02       &   2.00  $\pm$0.05    &0.65$\pm$0.05  &   0.3$\pm$0.1
\\
\hline
    Exp     & 1.50$\pm$0.05 &  2.0$\pm$0.2 & 0.57$\pm$0.10 & 0.30$\pm$0.10
\\
\hline \hline $\mu = 2$   &    $\tau$     & $\alpha$  &
$1/\sigma_k$  &   $\Delta_k$
\\
\hline
     RG     &     5/4      &   10/7   &      2/3      &      7/18
\\
\hline
    Sim     &     1.27$\pm$0.02       &   1.50 $\pm$0.05     &0.72$\pm$0.03  &
0.39$\pm$0.05\\ \hline
    Exp     & 1.27$\pm$0.03 & $1.5\pm$0.1 & 0.79$\pm$0.10 & 0.46$\pm$0.10
\\
\hline
\end{tabular}
\caption{The critical exponents measured in experiments compared
with RG results \protect\cite{NAT-92,NAR-93,LES-97,ERT-94} valid
to order $\epsilon\equiv 2\mu-d$ (see Ref.~\protect\cite{ZAP-98}
for the scaling relations) and simulations for the two
universality classes. System sizes range up to $L=150$ for $\mu =
1$ and to $L=400$ for $\mu = 2$.} \label{tab}
\end{table}

\begin{figure}[htb]
\narrowtext
\centerline{
        \epsfxsize=8.0cm
                \epsfbox{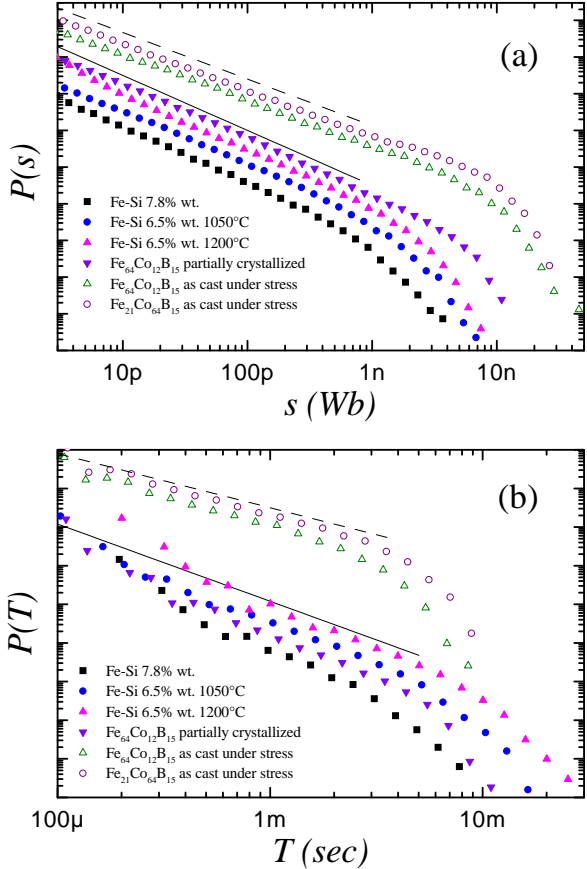}
        }
\caption{(a) Distributions of Barkhausen jump sizes measured in
different materials for the lowest available driving frequency.
The solid line has a slope $\tau = 1.5$ while for the dashed one
$\tau=1.27$, corresponding to the two universality classes; (b)
Similar plot for duration distributions. The solid line has a
slope $\alpha= 2$ while for the dashed one $\alpha =1.5$.}
\label{fig:pds}
\end{figure}
\begin{figure}[htb]
\narrowtext
\centerline{
        \epsfxsize=8.0cm
                \epsfbox{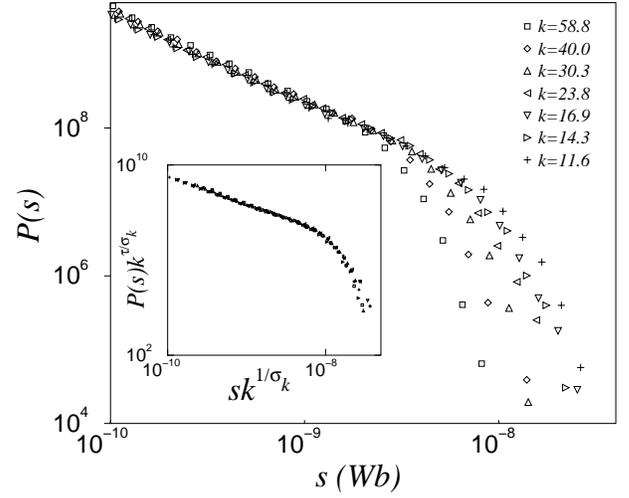}
        \vspace*{0.5cm}
        }
\caption{The Barkhausen jump size distribution for different
values of the demagnetizing factor $k$ in the
Fe$_{21}$Co$_{64}$B$_{15}$ amorphous alloy under tensile stress
The data collapse reported in the inset is done using
$\tau = 1.27$ and $1/\sigma_k = 0.78$.} \label{fig:psk}
\end{figure}
\begin{figure}[htb]
\narrowtext
\centerline{
        \epsfxsize=8.0cm
                \epsfbox{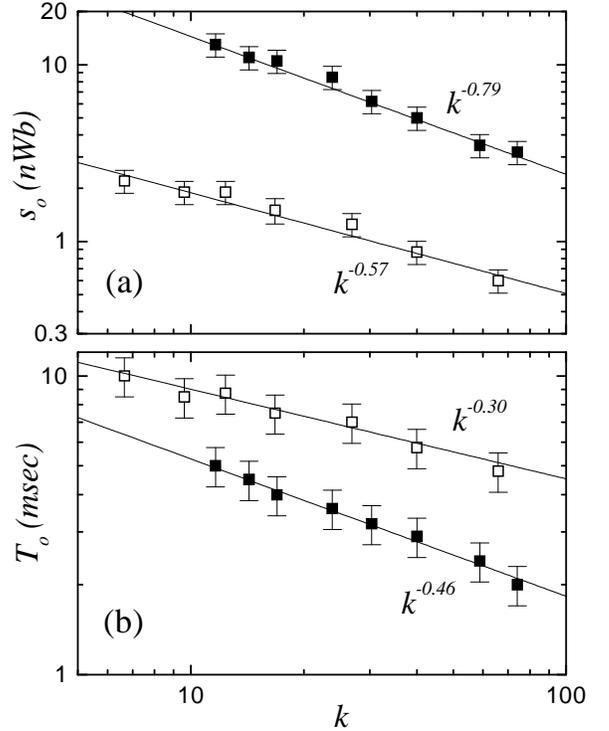}
        \vspace*{0.5cm}
        }
\caption{The cutoff of the Barkhausen size (a) and duration (b)
distributions as a function of $k$ in Fe-Si 1200$^\circ$C (empty
symbols) and Fe$_{64}$Co$_{21}$B$_{15}$ under constant tensile stress
(filled symbols).} \label{fig:s0t0}
\end{figure}

\end{multicols}

\begin{thebibliography}{10}
\bibitem{note_k}
We estimate the error bar for $k$ to be considerably smaller
than the error bars for $s_0$ and $T_0$.
\bibitem{Nota}
Contrary to the other exponents, $\nu_k$ does not receive
loop corrections \cite{NAR-96,NAR-93}.
\end{thebibliography}
\end{document}